\documentclass[twocolumn,showpacs,amsmath,amssymb,floatfix,superscriptaddress,aps,pre]{revtex4-1}
\pdfoutput=1
\usepackage{graphicx,color,subfigure} 
\usepackage{bm,bbold,empheq,mathdots}
\usepackage{units,enumitem}
\usepackage{hyperref} 
\usepackage{relsize}

 \newcommand{\be}{\begin{equation}}
 \newcommand{\ee}{\end{equation}}
  \newcommand{\xref}[1]{(\ref{#1})}

\begin{document}
\title{Tension-induced binding of semiflexible biopolymers
}

\author{Panayotis Benetatos}
\affiliation{Department of Physics, Kyungpook National University, 80 Daehak-ro, Buk-gu, Daegu, 702-701,  Korea}

\author{Alice von der Heydt}
\affiliation{Institut f\"ur Theoretische Physik, Georg-August-Universit\"at G\"ottingen, 
Friedrich-Hund-Platz 1, 37077 G\"ottingen, Germany}
\affiliation{Max-Planck-Institut f\"ur Dynamik und {Selbst}\-organisation, Am Fa{\ss}berg 17, 
37077 G\"ottingen, Germany}

\author{Annette Zippelius}
\affiliation{Institut f\"ur Theoretische Physik, Georg-August-Universit\"at G\"ottingen,
Friedrich-Hund-Platz 1, 37077 G\"ottingen, Germany}
\affiliation{Max-Planck-Institut f\"ur Dynamik und {Selbst}\-organisation, Am Fa{\ss}berg 17, 
37077 G\"ottingen, Germany}

\date{\today}

\begin{abstract}
  We investigate theoretically the effect of polymer tension on the
  collective behavior of reversibly binding cross-links.
  For this purpose, 
  we employ a
  model of two weakly bending wormlike chains aligned in parallel by a tensile force, 
   with a sequence of 
  inter-chain binding sites
  regularly spaced along the contours. 
  Reversible cross-links attach and detach at the sites with an
  affinity controlled by a chemical potential. In a mean-field
  approach, we calculate the free energy of the system and find the
  emergence of a free-energy barrier which controls the reversible
  (un)binding. The tension affects the conformational entropy of the
  chains which competes with the binding energy of the
  cross-links. This competition gives rise to a sudden increase in the
  fraction of bound sites as the tension increases. We show that this
  transition is related to the cross-over between weak and strong
  localization of a directed polymer in a pinning potential. The
  cross-over to the strongly bound state can be interpreted as a mechanism
  for force-stiffening of cross-linked polymers --- beyond the
  elasticity of a single wormlike chain.
\end{abstract}

\maketitle

\section{Introduction}

Cells and tissues are sensitive and responsive to mechanical forces.
The constituent 
filament networks are able to adapt 
in a differentiated manner
to a variety of strain or flow conditions, substrates, and
biological functions \cite{Ter14revSoftM,Picu11revSoftM}.  
Moreover, many cellular processes such as motion, adhesion, mitosis, and stress relaxation
require a reorganization of the cytoskeleton.
This remodeling can be achieved by transient or reversible cross-linkers
that bind and unbind 
stochastically with characteristic on- and off-rates. 
Reversible unbinding provides biopolymer networks with subtle relaxation mechanisms
and ultimately even allows them to flow \cite{Claus12NJP,Broedersz10PRL,Astrom2008}.
According to a model for the dynamics of the transient network \cite{Broedersz10PRL}, 
already a single time scale of cross-link unbinding 
gives rise to a broad spectrum  of relaxation times.
Experiments prove that thermal unbinding
of cross-linkers affects substantially the viscoelastic properties of actin networks,
and may result in
cross-link-induced stiffening~\cite{Lieleg10revSoftM,Gardel04Science}. Cytoskeletal filaments organize in branched or bundled structures with markedly different properties \cite{Schmoller2009}. The stress fibers are reversibly cross-linked actin bundles extending between focal adhesions which are the sites where cells adhere to the extracellular matrix. They have been the subject of significant experimental and modeling activity in recent years \cite{Colombelli,Lu2008,Hirata2008,Yoshigi2005,Burridge2013,Deguchi2006,Kumar2006,Machida2010}.

Apart from 
thermal unbinding, which in vitro can be controlled by temperature,
cross-link (un)binding prompted by mechanical forces 
is a potentially crucial mechanism to tune the fiber or network rigidity.
Substantial 
effort is made to reproduce mechanical stimuli on cytoskeletal filaments in the laboratory \cite{Eyckmans11}, to unravel, \textit{e.g.}, the signaling route to
stretch-induced reinforcement of actin stress fibers \cite{Hoffm12stress-fibers,Colombelli,Hirata2008,Yoshigi2005}.
Some of the experiments that highlight the importance of biomechanical signaling in stress fibers involve focused laser irradiation (laser nanoscissor surgery) \cite{Kumar2006,Yasukuni2007,Colombelli}, AFM probes \cite{Lu2008,Machida2010}, or substrate stretching \cite{Hirata2008,Yoshigi2005}.
Stretching forces 
can induce the adhesion of soft membranes (cells, vesicles) 
to a substrate~\cite{HelfrichBook,SeifertPRL95,SackmBruinsma02,SenguptaPRL10},
activate biochemical signals in the cytoskeleton~\cite{Tamada04},
and even change gene expression 
\cite{Wojtowicz10}.
For highly cross-linked microtubules, microrheological experiments have identified force-induced cross-link unbinding as the dominant contribution to viscoelasticity \cite{YangMicrorheo13}.
By studying a model of a reversibly bound actin filament bundle,
Heussinger et al.~\cite{Claus11coop-unbind-bundle,RichardClaus12} 
showed that unbinding is a cooperative effect, 
characterized by a free-energy barrier.

This study aims to shed light on a collective path 
to stress-induced fiber (network) 
stiffening via reversible binding, 
by means of a minimal model. 
We analyze 
two 
reversibly cross-linked semiflexible polymers
aligned in parallel by a longitudinal force.
For small tensile forces, we
expect that most cross-links are unbound, because constraining the transverse fluctuations of
the two filaments 
costs a substantial amount of entropy. 
Conversely, in a strongly stretched configuration, many cross-links are expected
to be in the bound state, because the cost of entropy is low.  
These expectations can be made quantitative within mean-field theory: 
The system is found to undergo a discontinuous phase transition from a
weakly bound state at low tension to a strongly bound state at high
tension. 

The melting (base-pair opening) transition of double-stranded DNA 
is also known both to impact the bending flexibility and to be affected by external forces \cite{GrossDNAmeltNatPhys11}.
Marenduzzo et al.~\cite{Marenduzzo09} consider DNA melting under stretching 
and determine the critical force as a function of temperature. 
In thermal equilibrium, so-called bubbles are present and a
central issue of theoretical studies \cite{TheoPeyr12}.
While our model bears some resemblance to and 
might be relevant for this transition,
one should keep in mind
that DNA is much more flexible than the cytoskeletal filaments, 
hence usually modeled as a self-avoiding random walk.
Furthermore, torsion is believed to be essential for double-stranded DNA.

This paper is organized as follows. In Sec.~\ref{sec:model}, we introduce the model of two parallel-stretched semiflexible chains with regularly spaced, reversible cross-links. 
In Sec.~\ref{sec:mean-field}, we introduce the average fraction of bound cross-links as an order parameter and calculate the free energy of our system in mean-field theory. 
The analysis of the free energy yields a mean-field first-order transition from a weakly bound to a strongly bound state as the stretching force or the binding affinity (controlled by the chemical potential) increases. 
In Sec.~\ref{sec:directed}, we discuss the cross-over between weak and strong localization of a directed polymer in a confining transverse potential well,
a system behaving similarly 
to that of the two cross-linked chains. 
Final remarks and conclusions are given in Sec.~\ref{sec:conclusion}.

\section{\label{sec:model}Model}

\begin{figure}
\includegraphics[width=\columnwidth]{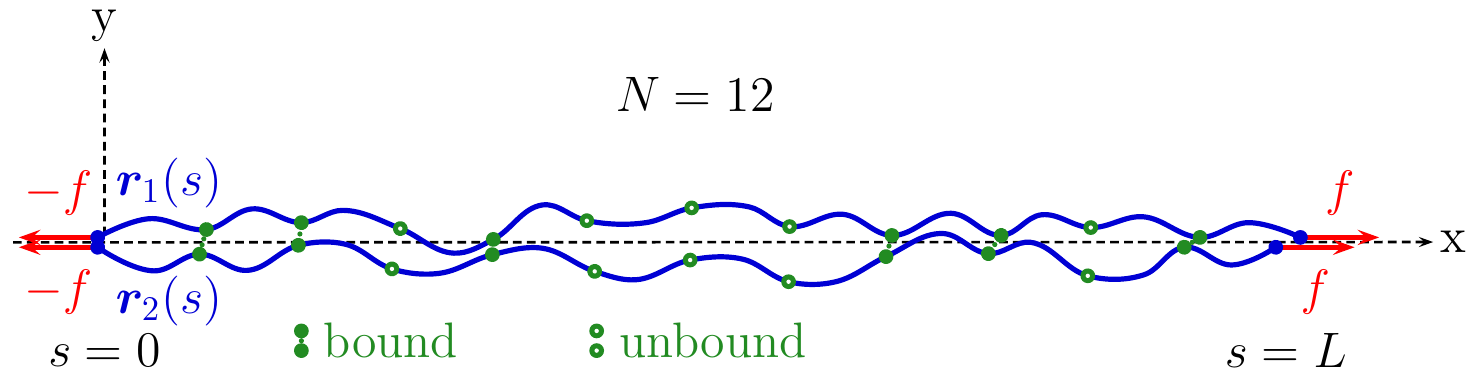}
\caption{\label{fig:model}Model sketch for $11$ reversible inter-chain cross-links.}
\end{figure}

The basis for our model are two identical semiflexible (inextensible) polymers which can reversibly bind to each other at equally spaced contour positions.
Both polymers have contour length $L$ 
and are aligned parallel along a given direction $x$ by a tensile force $f$, 
cf.\ Fig.~\ref{fig:model}. 
For simplicity, we consider two spatial
dimensions and use the weakly-bending approximation~\cite{MarkoSiggia1995}. 
This model is a direct generalization of that with permanent cross-links employed in \cite{AliceX-linkedWLC13}. 
The effective Hamiltonian (elastic-energy functional) is given in terms of
the transverse displacements $\left\{y_j(s)\right\}$, $s\in[0,L]$, $j=1,2$, by
\begin{align}
{\cal{H}} = 
&
\underbrace{
\sum_{j=1}^2 \int_0^L ds 
\left(
\frac{\kappa}{2}  \left( \partial^2_{s} y_j \right)^2
+ \frac{f}{2} \left( \partial_{s} y_j \right)^2
\right)
- 2fL
}_{{\cal H}_0}
\notag\\
& \mbox{} + \frac{g}{2} \sum_{b=1}^{N-1} n_b\Bigl( y_1(bd) - y_2(bd) \Bigr)^2
\label{H-total}
\end{align}
where $\kappa$ is the bending rigidity,
${\cal H}_0$ is the Hamiltonian of the two polymers without
cross-links, and the last term accounts for the cross-links at contour sites $bd$ 
of spacing $d\mathrel{\mathop:}=L/N$. 
If cross-link $b$ is bound, it acts as an harmonic spring of strength $g$
and the binary variable $n_b$ assumes the
value $1$, 
otherwise $n_b = 0$.
We assume hinged-hinged boundary conditions, 
which suppress transverse displacements and curvatures
at the end points, $y_j(0)=y_j(L)=0$ and $\partial^2_{s} y_j(0) = \partial^2_{s} y_j(L) = 0$, for $j=1,2$. In addition, we impose a no-slip condition, $x_1(0)=x_2(0)$.
The total number of cross-links fluctuates controlled by a
chemical potential $\mu$, so that the grand-canonical partition function reads
\be
\mathcal{Z} =
\int\! {\cal D}[y_1,y_2] 
\prod_{b=1}^{N-1}\left( \sum_{n_b=0}^1\right)
\text{e}^{ -\beta \left( {\cal H} - \mu\sum_{b=1}^{N-1} n_b \right) },
\ee
where the functional integral $\int\!{\cal D}[y_1,y_2]$ comprises all polymer conformations consistent with the boundary conditions, and $\beta \mathrel{\mathop:}= 1/(k_{\text B} T)$.



\section{\label{sec:mean-field}Mean-field treatment of many reversible cross-links}


Following a mean-field approach analogous to \cite{Claus11coop-unbind-bundle}, 
we replace the
individual cross-link degrees of freedom, $n_b$, with their average value
\be 
n \mathrel{\mathop:}= \frac{\tilde N}{N-1} \mathrel{\mathop:}=
\frac{1}{N-1}\sum_{b=1}^{N-1}n_b, 
\ee 
$\tilde N=n(N-1)$ denoting the total number of bound cross-links.
This way, we find for
the grand-canonical partition function
in mean-field approximation, relative to that of the polymers without binding sites, 
\be\label{mfpart-fun} 
\mathcal{Z}_{\text{mf}} = \sum_{{\tilde
    N}=0}^{N-1} \binom{N-1 
    }{\tilde N} \exp(\beta \mu {\tilde N})
\mathcal{Z}_{N-1}(ng).
\ee 
Here, $\mathcal{Z}_{N-1}(ng)$ is the relative canonical partition function of two weakly bending chains with exactly $N-1$
\textit{irreversible\/} crosslinks 
of effective strength $ng$ [substitute $n_b\rightarrow n$  in Eq.~\xref{H-total}]. 
This partition function has been computed exactly in \cite{AliceX-linkedWLC13} and is given by
\begin{align}\label{eq-Z_b}
\mathcal{Z}_{N-1}(ng)
&=\prod_{l=1}^{N-1} Z_{l}(n),\nonumber\\ 
 Z_{l}(n)
& = \left\{  1 + \frac{n g d}{f} \Bigl( \psi_{l}(0) - \psi_{l} (\delta_f) \Bigr) \right\}^{-1/2},\\ 
\nonumber
\psi_{l}(\delta_f) 
& = \frac{\sinh \delta_f}{\delta_f (\cosh \delta_f - \cos \varphi_{l})},\\ \nonumber
\end{align}
with the cross-link spacing $d=L/N$, 
the force parameter
\be
\delta_f:=d\sqrt{f/\kappa},
\ee 
and $\varphi_l:=\pi l/N$.
Using the Stirling approximation for $N\gg {\tilde N} \gg 1$, we obtain
for the mean-field free energy per cross-link site, $(N-1)\beta G_{\text{mf}} \mathrel{\mathop:}= - \ln {\cal Z}_{\text{mf}}$,
\begin{align}
\label{mffree_en}
\beta G_{\text{mf}} =\, & n \ln n + (1-n) \ln(1-n) - \beta \mu n
\notag\\ & \mbox{} +\frac{1}{2(N-1)}\sum_{l=1}^{N-1}\ln Z_{l}(n).
\end{align}

\begin{figure}
\includegraphics[width=\columnwidth]{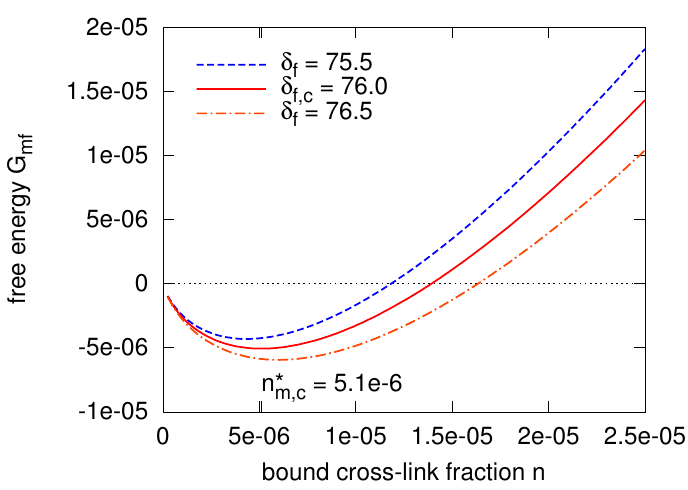}
\caption{\label{small_n}%
  Mean-field free energy $G_{\text{mf}}(n)$ for very small n; parameters are
  chosen as $N=20$ (19 possible binding sites), $\beta\mu=-0.1$,
  cross-link strength $w=10^4$ and 3 different force parameters
  $\delta_f$; a minimum for very small $n$, corresponding to a weakly
  bound state, exists for all $\delta_f$.}
\end{figure}

The equilibrium value of $n$, which acts as the order parameter for the
binding-unbinding transition, is the one that minimizes the free
energy $G_{\text{mf}}$ 
at fixed values of the other parameters of our model.
These are: 
the effective force strength $\delta_f^2=d^2 f/\kappa$, the effective cross-link
strength $2w \mathrel{\mathop:}= g d^3/\kappa$ (in Eq.~\xref{eq-Z_b}, $gd/f = 2w/\delta_f^2$), the chemical potential $\beta\mu$; 
the systems under consideration are additionally parameterized by
the number of cross-link sites, $N-1$. 
At least for finite $N$, the behavior of $G_{\text{mf}}(n)$ for
very small $n$ is dominated by the entropic contribution $n\ln(n)$,
implying a negative slope of $G_{\text{mf}}(n)$ for very small $n$ 
independently of the choice of the other parameters,
and hence a weakly bound state with a very small fraction of bound cross-links. 
A blow-up of this region is shown in Fig.~\ref{small_n}, where $n_{m,c}^*$ denotes
the bound cross-link fraction at this free-energy minimum, whose value itself 
is extremely close to (below) zero. 
\begin{figure}
\includegraphics[width=\columnwidth]{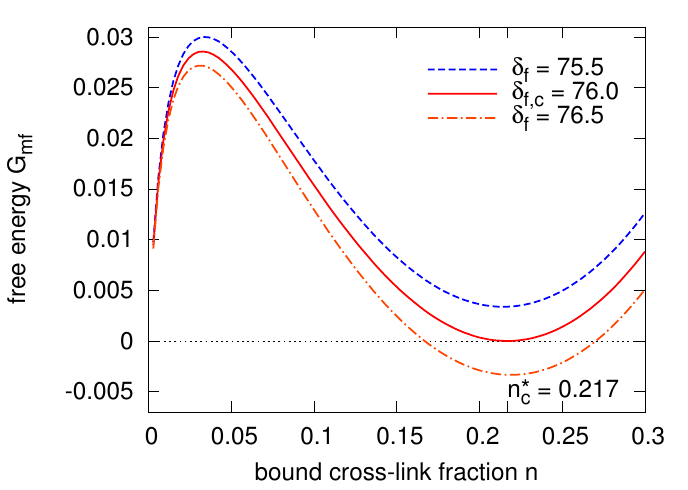}
\caption{\label{G-n}%
  Free energy $G_{\text{mf}}(n)$; the second
  minimum becomes globally stable at $\delta_{f,c}=76.0$; 
  parameters as in Fig.~\ref{small_n}.}
\end{figure}
Whether or not a markedly bound state with a substantial fraction of bound cross-links
exists in mean-field theory, depends on the parameters of our model. 
In Fig.~\ref{G-n},
we show $\beta G_{\text{mf}}(n)$ over a larger range of $n$-values, revealing the
formation of a second minimum of $G_{\text{mf}}$ at a value of $n$
considerably larger than zero.
We can locate the transition  to the markedly or strongly bound state by computing for each given set of parameters the fraction $n^* > n_{m,c}^*$ 
which minimizes $G_{\text{mf}}$ (with $G_{\text{mf}}(n^*) \geq 0$), 
and vary the parameters until $n^*=n_c^*$
additionally fulfills $G_{\text{mf}}(n_c^*) = 0$.
Thus, the transition is first order in mean-field theory with a
discontinuous jump in the order parameter from a very small to a sensibly large
value between zero and one. 

\begin{figure}
\includegraphics[width=\columnwidth]{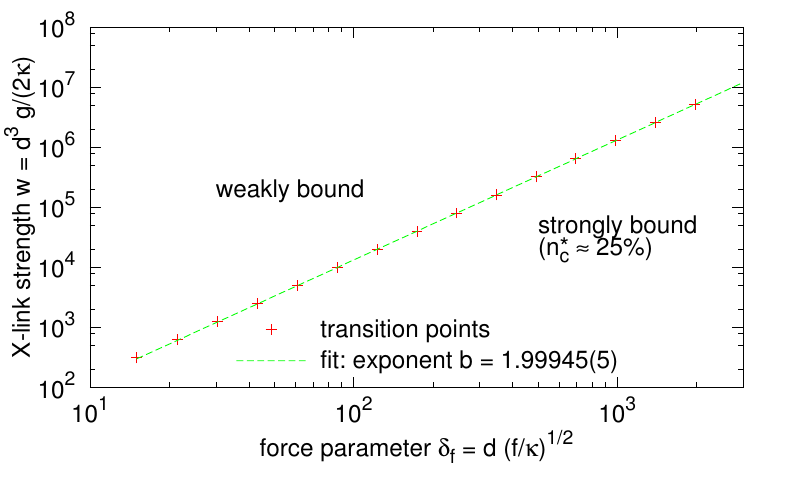}
\caption{Phase diagram in the plane of effective force and cross-link strength; 
symbols: first-order transition data obtained from mean-field theory; 
dashed line: fit with scaling relation derived from model of directed polymer in a pinning potential (Sec.~\ref{sec:directed}); 
the chosen parameters are $N=300$ and $\beta\mu=-0.1$.
  \label{g_deltaf}}
\end{figure}

\begin{figure}
\includegraphics[width=\columnwidth]{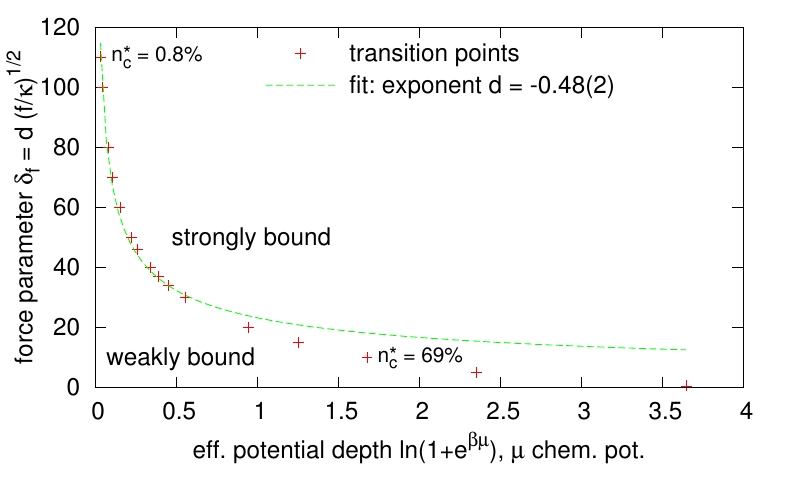}
\caption{Phase diagram in the plane of effective force and chemical potential; symbols: first-order transition obtained from mean-field theory; dashed line: fit with scaling relation from model of directed polymer in a pinning potential (Sec.~\ref{sec:directed}); parameters are $N=300$ and $w=1000$.
\label{deltaf_mu}}
\end{figure}

Locating the binding-unbinding transition allows us to map out the phase diagram in several parameter planes, 
\textit{e.g.\/}, in the plane spanned
by force and cross-link strength, or in the plane of force
and chemical potential. 
The cross-link strength against force parameter at the transition 
is shown in Fig.~\ref{g_deltaf} (symbols indicate data obtained by minimization of
$G_{\text{mf}}$). 
For all parameter values, we clearly observe 
a sharp transition between a weakly bound and a strongly bound state. 
As the force (or $\delta_f$) is increased, the bound state is increasingly preferred, 
because the entropy loss due to cross-linking is reduced. 
Increasing $g$ (or $w$) implies a decrease in the range of the
cross-link potential, hence a stronger localization and 
a more pronounced reduction of entropy for each bound cross-link, 
giving rise to a larger region of weakly bound states. 
Instead of increasing the force $f$, we can alternatively increase the
chemical potential $\beta\mu$ (the affinity for binding) to control the
transition from the weakly bound to the strongly bound state.  
The phase diagram in the plane of $f$ and $\beta\mu$ is shown in Fig.~\ref{deltaf_mu}.
As expected, the transition occurs at smaller $f$, if the affinity for cross-link binding is large. Furthermore, the fraction $n_c^*$ of bound cross-links 
increases monotonically with $\beta\mu$ as indicated by two values in
Fig.~\ref{deltaf_mu}. 
Beyond these qualitative considerations, it is actually possible to derive scaling
relations for the transition parameters, 
the fits to which are represented as dashed lines in Figs.\ \ref{g_deltaf} and \ref{deltaf_mu}.  
The scaling relations will be derived in Sec.~\ref{sec:directed}.

\begin{figure}
\includegraphics[width=\columnwidth]{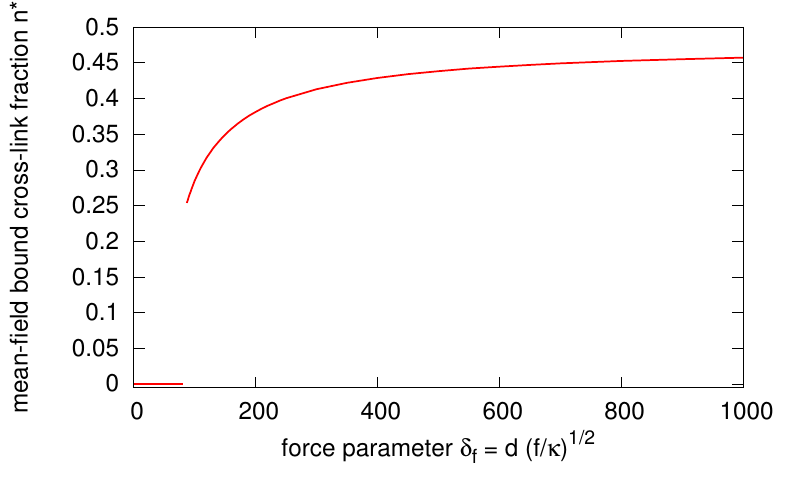}
\caption{Fraction of bound cross-links as a function of force for parameters $N=500$, $\beta\mu=-0.1$ and $w=10^4$.
\label{fraction}}
\end{figure}

Eventually, Fig.~\ref{fraction} visualizes the jump of the order parameter, 
the mean-field fraction of bound cross-links,
as a function of stretching force (the weakly bound state cannot be resolved on this scale).

\section{Directed polymer in a transverse potential well}\label{sec:directed}

In this section, we shall discuss a related model of a single
directed polymer in a continuous well potential. First, we give a
plausibility argument, why this model should be comparable to our system of two
reversibly cross-linked polymers. Then, we briefly review the main
features of the model, which has been analyzed \cite{NelsonBook} in the
context of magnetic flux lines in a high-$T_c$ superconductor being
pinned by columnar defects. Transferring these results to our model 
allows us to derive and test scaling predictions for
the transition lines in the phase diagram (Sec.~\ref{scal-rel}).

First, we
observe that reversible cross-links provide an effective, short-range attractive
potential for the two connected polymers. 
This is most easily seen by considering a single cross-link site. 
Integrating out the cross-link degree of freedom,
we obtain an effective potential \cite{Kierfeld05} as a function of
the inter-polymer distance $\Delta y\mathrel{\mathop:} = y_1(L/2) - y_2(L/2)$ 
at the cross-link site,
\be
\label{eff-pot}
\beta V_{\text{eff}}(\Delta y) = -\ln\left[
1 + \exp\left\{
\beta\left( \mu 
- \frac{g}{2} \Delta y^2 
\right)\right\} 
\right],
\ee
\begin{figure}
\includegraphics[]{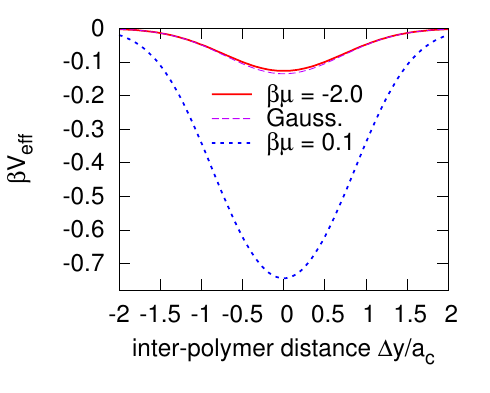}
\caption{\label{fig:V_eff}Effective cross-link potential as a function of the inter-polymer distance $\Delta y$.}
\end{figure} 
shown in Fig.~\ref{fig:V_eff} for two representative values of $\beta\mu$.  

Assuming $\mu<0$, the effective potential is approximated by
an inverted Gaussian potential
\be\label{eff-pot1}
\beta V_{\text{eff}} (\Delta y)
\approx 
- \text{e}^{\beta \mu}
\exp\left(
- \frac{\Delta y^2}{a_c^2}
\right),
\ee
whose range is $a_c=\sqrt{2 k_{\text B} T/g}$ and whose amplitude or depth is $U_0 \mathrel{\mathop:}= k_{\text B} T \exp(\beta \mu)$ [in general, the depth is $k_{\text B} T\ln(1+\text{e}^{\beta\mu})$].
To further ease the setup of the analogy with a flexible directed polymer 
in a square well,
one may
approximate the effective potential by
$V_{\text{eff}}(\Delta y)=-U_0$ for $-a_c<\Delta y<a_c$ and $V_{\text{eff}}(\Delta y)=0$ otherwise. 
Furthermore, for many equally
spaced cross-link sites, the effective potential acts approximately
continuously along the polymer length --- 
like the potential well extending continuously in the direction of the tensile force.

Finally, we assume that the long wavelength transverse excitations are dominated 
by the second term of $\mathcal{H}_0$ in 
Eq.~\xref{H-total}, so that we can treat one of the polymers as a 
flexible, \textit{directed\/} polymer in a short-range potential well that constrains the
transverse displacements. 
This assumption holds in the strong-stretching limit, for which the ``memory'' length $l_m \mathrel{\mathop:}=\sqrt{\kappa/f}$ is much smaller than the site spacing $d$, $l_m\ll d$. 
In this limit, the off-boundary, transverse mean-square fluctuations of a stretched semiflexible polymer 
are known to scale as $\sim k_{\text B} T L/f$, \textit{i.e.\/}, to become independent of the bending rigidity $\kappa$ \cite{PBEMT10R}. 

The partition function of a flexible
directed chain, subject to a tensile force with boundary condition $y(0)=y(L)=0$ and to a well potential $V(y)$ continuous in tension direction, 
relative to the free (noninteracting) chain, is given by the path integral
\begin{widetext}
\begin{align}
\label{pathintegral}
{\cal Z}_{\text{rel}}=\frac{\mathlarger{\int_{y(0)=0}^{y(L)=0}}{\cal D}y(s) \exp\Big[-\mathlarger{\frac{\beta f}{2}}\mathlarger{\int_0^L}\mathlarger{\Big(\frac{dy}{ds}\Big)^2}ds-\beta\mathlarger{\int_0^L}V[y(s)]ds\Big]}{\mathlarger{\int_{y(0)=0}^{y(L)=0}}{\cal D}y(s) \exp\Big[-\mathlarger{\frac{\beta f}{2}}\mathlarger{\int_0^L}\mathlarger{\Big(\frac{dy}{ds}\Big)^2}ds\Big]}.
\end{align}
\end{widetext}
When the polymer is confined within the  well, there is a competition
between energy gain and
entropy loss, the latter depending on the tensile force on the chain.

We notice that the
path integrals in the numerator and the denominator correspond to
density matrix elements of a fictitious quantum particle in a
potential $V(y)$ and a free one, respectively \cite{Feynman}. 
The mapping is as follows: $f \leftrightarrow m$, $\beta \leftrightarrow
\hbar^{-1}$, $s \leftrightarrow t$, $L \leftrightarrow {\tilde \beta}
\hbar$, where $m$, $t$, ${\tilde \beta}$ are mass, time, and
inverse temperature parameter ($1/k_B{\tilde T}$) for the fictitious
quantum particle. The thermodynamic limit ($L\rightarrow \infty $) for
the polymer corresponds to the zero temperature limit (${\tilde
  \beta}\rightarrow \infty$) for the fictitious quantum particle.
  In this limit, the density matrix of a quantum particle is dominated by
the ground state:
\begin{align}
\rho(y,y';{\tilde \beta})\approx \psi_0(y)\psi_0^*(y')\exp(-{\tilde \beta}E_0),
\end{align}
where $\psi_0(y)$ is the ground-state eigenfunction and $E_0$ the ground-state energy.
Using the aforementioned correspondence, we obtain:
\begin{align}\label{Z-rel-dir}
{\cal Z}_{\text{rel}}=\Big(\frac{2\pi L k_B T}{f}\Big)^{1/2}\psi_0(0)\psi_0(0)\exp(-\beta LE_0),
\end{align}
where the prefactor arises from the density matrix of the free
particle. In the limit $L\to\infty$, the binding free energy per unit
length can be extracted from Eq.~\xref{Z-rel-dir} using ${\cal
  Z}_{\text{rel}}=\exp(-\beta L G)$, and we obtain $G=E_0$.

We point out that in $(1+1)$ dimensions, as well as in $(1+2)$
dimensions, the fictitious quantum particle always has a bound state,
implying that the corresponding directed polymer is always
bound, in agreement with our data set presented in Fig.~\ref
{small_n}.  However, there is a clear cross-over between a strongly
bound and a weakly bound state, and it is this cross-over which is
captured by the mean-field analysis.

For a strong stretching force, the entropic contribution to the free
energy is small compared to the energy of the potential well which can
be treated as infinitely deep. In this case, the corresponding quantum
problem is that of a particle in an infinitely deep well. From
elementary quantum mechanics \cite{Landau}, we obtain the strong-confinement free energy per unit length
\begin{align}
G_{s} = -U_0 + \frac{\pi^2}{4}\frac{(k_BT)^2}{2  f a_c^2}.
\end{align}
The second term in the rhs of the above equation is the entropy loss
due to the confinement in the effective potential, which in the strong-stretching case
is much smaller than the first term. 
In effect, the particle is localized within the width of the effective potential well,
$a_c$. 

In the opposite limit of weak localization, the stretching
force is smaller and allows the polymer to perform fluctuations far beyond the range
of the potential well, which can be treated as a delta-function:
$V_1(y)=-U_0 2 a_c \delta(y)$.  Using the ground state of the
corresponding quantum problem, the weak-confinement free energy per
unit length becomes
\begin{align}
G_w=-\frac{f}{2 (k_bT)^2}(U_0 2 a_c)^2.
\end{align}
In this case, it is known that the wave function of the fictitious
quantum particle is $\propto \exp(-|y|/l_{\perp})$, with
$l_{\perp}=(k_B T)^2/(fU_0 2 a_c)$.

Hence,
the crossover between strong and weak binding occurs at a force $f_c$ such that
\begin{align}\label{rel-crit-par}
  \frac{(k_BT)^2}{2  f_c a_c^2}=U_0.
\end{align}

\subsection{\label{scal-rel}Scaling predictions for our system}

Exploiting the analogy of our system with a directed polymer in a square well,
we can extract from Eq.~\xref{rel-crit-par} 
scaling relations between the parameters at the binding-unbinding transition (for transparency, we omit the label $c$ here). 
The force $f$ ($\propto \delta_{f}^2$ at constant site spacing $d$, bending rigidity $\kappa$, and $\beta$) is expected to increase linearly with the inverse squared potential width $1/a_c^2$ ($\propto w$ at constant $d$, $\kappa$, and $\beta$), \textit{i.e.\/}
\be
\delta_{f}^2 \propto w.
\ee 
Inspection of Fig.~\ref{deltaf_mu} shows that
the computed transition obeys this scaling relation with a high precision.
In fact, this scaling should be inherent to our model, since the cross-link strength enters the free energy, Eq.~\xref{mffree_en}, only via the ratio $w/\delta_f^2$.
On the basis of Fig.~\ref{deltaf_mu}, we test the expected linear scaling of the force with the inverse potential depth $1/U_0$ at the transition,
cf.\ the parameter mapping following Eq.~\xref{eff-pot}, viz.\
\be
\delta_{f}^2 \propto (\beta U_{0})^{-1} = \left[ \ln( 1 + \text{e}^{\beta\mu}) \right]^{-1}.
\ee
We notice that the scaling of force with potential depth is reasonably good, 
but not as convincing as that  with potential width.
One reason is that the points which are markedly off the scaling curve correspond to small stretching forces, for which the analogy with the (flexible) directed polymer breaks down. 
Another reason appears to be that particularly for positive $\beta\mu$, 
identifying the square-well depth with the minimum of the 
parabola-shaped, effective potential, cf.\ Fig.~\ref{fig:V_eff}, is a less accurate approximation.

\section{\label{sec:conclusion}Conclusions}

We have studied two reversibly cross-linked, semiflexible filaments
under tension and shown that the two filaments are always in a
(weakly) bound state. Within mean-field theory, there is a
discontinuous phase transition as a function of applied stretching
force from a weakly bound state at small force to a strongly bound state
at large force. The critical force as a function of either cross-link
strength or chemical potential displays a scaling, 
which can be derived from a model of a single polymer in a confining potential. 
Our analysis has been restricted to two dimensions,
but neglecting entanglement effects and twist, 
our results can be generalized to $(1+2)$ dimensions, 
since in the weakly-bending approximation, 
the two transverse directions decouple.

The sharp transition obtained in mean-field theory
is expected to be replaced by a cross-over, if fluctuations are taken into account --
similar to the cross-over observed in the model of a single stretched polymer in
a confining potential. 
Nonetheless, the tension-induced binding cross-over from a weakly to a strongly bound state
predicted by our model
can be interpreted as a versatile
microscopic mechanism of force-stiffening.
It provides an extra contribution to supramolecular nonlinear elasticity insofar as
reversibly binding polymers are enabled
to tune their mechanical properties 
beyond those of a single semiflexible polymer: 
A larger tensile force causes more cross-links to bind, 
and a larger number of bound cross-links
results in a larger tensile stiffness of the system~\cite{AliceX-linkedWLC13}.
Thereby, the polymer pair responds to an increasing external force with a larger modulus 
to diminish the effect of this force. The main experimental signature that arises from our theoretical results would be a cross-link regulated tension stiffening of semiflexible filament bundles accompanied by an increase in the number of bound cross-links. In the model of Ref. \cite{Claus12NJP}, softening of the cytoskeletal network due to cross-link unbinding (due to bending)  competes with inherent stiffening due to a single filament's or cross-link's nonlinear elasticity.
It is an interesting open question to be clarified further by experiments
whether cells make use of the stiffening predicted by our model to adapt the mechanical properties of the cytoskeleton to the temporary task or substrate.

The described mechanism might also give clues to the formation and stress-induced strengthening of stress fibers. It is known that stretching induces mobilization of the protein zyxin from focal adhesions to  actin filaments and the stress fibers become thicker in a zyxin-dependent manner \cite{Yoshigi2005}. It is worth investigating whether this stress-fiber strengthening is actually a collective binding of reversible cross-links similar to that described in our model. Single stress-fiber experiments similar to those described in \cite{Deguchi2006} might be useful in this direction.

This study could be extended in several directions, some of which are work in progress.
First, it would be desirable to find a tractable description to compute not only the mean-field fraction of bound cross-links, but also the spatial correlation of bound cross-link sites. Longitudinal tension and boundary conditions imply that
transient cross-links are more likely to bind close to the polymer ends, 
an effect which might weaken the binding transition we spotted here. 
Second, the thermodynamic limit of continuous binding sites and its impact on the cross-over remains to be analyzed further.
An obvious step following the observation of the binding cross-over would be to compute the force-extension relation for transient binding and compare it to the irreversibly cross-linked case \cite{AliceX-linkedWLC13}.
The stretching elasticity of bundles of many parallel-aligned semiflexible polymers \cite{PanayotisStephanAnnette12,Claus07PRL} 
with reversible cross-links is another challenging problem. 
Finally, semiflexible polymers with more complicated binding modes, 
such as cross-links aligning (welding)  the corresponding filament segments in the stretching direction \cite{PBEFrey03,PBAZPRL07} or nonlocal binding \cite{LiverpoolEPL09}, 
should be investigated  
in the case of reversible cross-links.

\acknowledgments

We gratefully acknowledge support by the Deut\-sche For\-schungs\-ge\-mein\-schaft (DFG) via grant SFB-937/A1. Also, we thank Claus Heussinger for useful discussions.

\bibliography{reversible-binding}

\begin{thebibliography}{43}%
\makeatletter
\providecommand \@ifxundefined [1]{%
 \@ifx{#1\undefined}
}%
\providecommand \@ifnum [1]{%
 \ifnum #1\expandafter \@firstoftwo
 \else \expandafter \@secondoftwo
 \fi
}%
\providecommand \@ifx [1]{%
 \ifx #1\expandafter \@firstoftwo
 \else \expandafter \@secondoftwo
 \fi
}%
\providecommand \natexlab [1]{#1}%
\providecommand \enquote  [1]{``#1''}%
\providecommand \bibnamefont  [1]{#1}%
\providecommand \bibfnamefont [1]{#1}%
\providecommand \citenamefont [1]{#1}%
\providecommand \href@noop [0]{\@secondoftwo}%
\providecommand \href [0]{\begingroup \@sanitize@url \@href}%
\providecommand \@href[1]{\@@startlink{#1}\@@href}%
\providecommand \@@href[1]{\endgroup#1\@@endlink}%
\providecommand \@sanitize@url [0]{\catcode `\\12\catcode `\$12\catcode
  `\&12\catcode `\#12\catcode `\^12\catcode `\_12\catcode `\%12\relax}%
\providecommand \@@startlink[1]{}%
\providecommand \@@endlink[0]{}%
\providecommand \url  [0]{\begingroup\@sanitize@url \@url }%
\providecommand \@url [1]{\endgroup\@href {#1}{\urlprefix }}%
\providecommand \urlprefix  [0]{URL }%
\providecommand \Eprint [0]{\href }%
\providecommand \doibase [0]{http://dx.doi.org/}%
\providecommand \selectlanguage [0]{\@gobble}%
\providecommand \bibinfo  [0]{\@secondoftwo}%
\providecommand \bibfield  [0]{\@secondoftwo}%
\providecommand \translation [1]{[#1]}%
\providecommand \BibitemOpen [0]{}%
\providecommand \bibitemStop [0]{}%
\providecommand \bibitemNoStop [0]{.\EOS\space}%
\providecommand \EOS [0]{\spacefactor3000\relax}%
\providecommand \BibitemShut  [1]{\csname bibitem#1\endcsname}%
\let\auto@bib@innerbib\@empty
\bibitem [{\citenamefont {Pritchard}\ \emph {et~al.}(2014)\citenamefont
  {Pritchard}, \citenamefont {Huang},\ and\ \citenamefont
  {Terentjev}}]{Ter14revSoftM}%
  \BibitemOpen
  \bibfield  {author} {\bibinfo {author} {\bibfnamefont {R.~H.}\ \bibnamefont
  {Pritchard}}, \bibinfo {author} {\bibfnamefont {Y.~Y.~S.}\ \bibnamefont
  {Huang}}, \ and\ \bibinfo {author} {\bibfnamefont {E.~M.}\ \bibnamefont
  {Terentjev}},\ }\href@noop {} {\bibfield  {journal} {\bibinfo  {journal}
  {Soft Matter}\ }\textbf {\bibinfo {volume} {10}},\ \bibinfo {pages} {1864}
  (\bibinfo {year} {2014})}\BibitemShut {NoStop}%
\bibitem [{\citenamefont {Picu}(2011)}]{Picu11revSoftM}%
  \BibitemOpen
  \bibfield  {author} {\bibinfo {author} {\bibfnamefont {R.~C.}\ \bibnamefont
  {Picu}},\ }\href@noop {} {\bibfield  {journal} {\bibinfo  {journal} {Soft
  Matter}\ }\textbf {\bibinfo {volume} {7}},\ \bibinfo {pages} {6768} (\bibinfo
  {year} {2011})}\BibitemShut {NoStop}%
\bibitem [{\citenamefont {Heussinger}(2012)}]{Claus12NJP}%
  \BibitemOpen
  \bibfield  {author} {\bibinfo {author} {\bibfnamefont {C.}~\bibnamefont
  {Heussinger}},\ }\href@noop {} {\bibfield  {journal} {\bibinfo  {journal}
  {New J.\ Phys.}\ }\textbf {\bibinfo {volume} {14}},\ \bibinfo {pages}
  {095029} (\bibinfo {year} {2012})}\BibitemShut {NoStop}%
\bibitem [{\citenamefont {Broedersz}\ \emph {et~al.}(2010)\citenamefont
  {Broedersz}, \citenamefont {Depken}, \citenamefont {Yao}, \citenamefont
  {Pollak}, \citenamefont {Weitz},\ and\ \citenamefont
  {MacKintosh}}]{Broedersz10PRL}%
  \BibitemOpen
  \bibfield  {author} {\bibinfo {author} {\bibfnamefont {C.~P.}\ \bibnamefont
  {Broedersz}}, \bibinfo {author} {\bibfnamefont {M.}~\bibnamefont {Depken}},
  \bibinfo {author} {\bibfnamefont {N.~Y.}\ \bibnamefont {Yao}}, \bibinfo
  {author} {\bibfnamefont {M.~R.}\ \bibnamefont {Pollak}}, \bibinfo {author}
  {\bibfnamefont {D.~A.}\ \bibnamefont {Weitz}}, \ and\ \bibinfo {author}
  {\bibfnamefont {F.~C.}\ \bibnamefont {MacKintosh}},\ }\href@noop {}
  {\bibfield  {journal} {\bibinfo  {journal} {Phys.\ Rev.\ Lett.}\ }\textbf
  {\bibinfo {volume} {105}},\ \bibinfo {pages} {238101} (\bibinfo {year}
  {2010})}\BibitemShut {NoStop}%
\bibitem [{\citenamefont {{\AA}str{\"o}m}\ \emph {et~al.}(2008)\citenamefont
  {{\AA}str{\"o}m}, \citenamefont {Kumar}, \citenamefont {Vattulainen},\ and\
  \citenamefont {Karttunen}}]{Astrom2008}%
  \BibitemOpen
  \bibfield  {author} {\bibinfo {author} {\bibfnamefont {J.~A.}\ \bibnamefont
  {{\AA}str{\"o}m}}, \bibinfo {author} {\bibfnamefont {P.~B.~S.}\ \bibnamefont
  {Kumar}}, \bibinfo {author} {\bibfnamefont {I.}~\bibnamefont {Vattulainen}},
  \ and\ \bibinfo {author} {\bibfnamefont {M.}~\bibnamefont {Karttunen}},\
  }\href@noop {} {\bibfield  {journal} {\bibinfo  {journal} {Phys.\ Rev.\ E}\
  }\textbf {\bibinfo {volume} {77}},\ \bibinfo {pages} {051913} (\bibinfo
  {year} {2008})}\BibitemShut {NoStop}%
\bibitem [{\citenamefont {Lieleg}\ \emph {et~al.}(2010)\citenamefont {Lieleg},
  \citenamefont {Claessens},\ and\ \citenamefont {Bausch}}]{Lieleg10revSoftM}%
  \BibitemOpen
  \bibfield  {author} {\bibinfo {author} {\bibfnamefont {O.}~\bibnamefont
  {Lieleg}}, \bibinfo {author} {\bibfnamefont {M.~M.~A.~E.}\ \bibnamefont
  {Claessens}}, \ and\ \bibinfo {author} {\bibfnamefont {A.~R.}\ \bibnamefont
  {Bausch}},\ }\href@noop {} {\bibfield  {journal} {\bibinfo  {journal} {Soft
  Matter}\ }\textbf {\bibinfo {volume} {6}},\ \bibinfo {pages} {218} (\bibinfo
  {year} {2010})}\BibitemShut {NoStop}%
\bibitem [{\citenamefont {Gardel}\ \emph {et~al.}(2004)\citenamefont {Gardel},
  \citenamefont {Shin}, \citenamefont {MacKintosh}, \citenamefont {Mahadevan},
  \citenamefont {Matsudaira},\ and\ \citenamefont {Weitz}}]{Gardel04Science}%
  \BibitemOpen
  \bibfield  {author} {\bibinfo {author} {\bibfnamefont {M.~L.}\ \bibnamefont
  {Gardel}}, \bibinfo {author} {\bibfnamefont {J.~H.}\ \bibnamefont {Shin}},
  \bibinfo {author} {\bibfnamefont {F.~C.}\ \bibnamefont {MacKintosh}},
  \bibinfo {author} {\bibfnamefont {L.}~\bibnamefont {Mahadevan}}, \bibinfo
  {author} {\bibfnamefont {P.}~\bibnamefont {Matsudaira}}, \ and\ \bibinfo
  {author} {\bibfnamefont {D.~A.}\ \bibnamefont {Weitz}},\ }\href@noop {}
  {\bibfield  {journal} {\bibinfo  {journal} {Science}\ }\textbf {\bibinfo
  {volume} {304}},\ \bibinfo {pages} {1301} (\bibinfo {year}
  {2004})}\BibitemShut {NoStop}%
\bibitem [{\citenamefont {Schmoller}\ \emph {et~al.}(2009)\citenamefont
  {Schmoller}, \citenamefont {Lieleg},\ and\ \citenamefont
  {Bausch}}]{Schmoller2009}%
  \BibitemOpen
  \bibfield  {author} {\bibinfo {author} {\bibfnamefont {K.~M.}\ \bibnamefont
  {Schmoller}}, \bibinfo {author} {\bibfnamefont {O.}~\bibnamefont {Lieleg}}, \
  and\ \bibinfo {author} {\bibfnamefont {A.~R.}\ \bibnamefont {Bausch}},\
  }\href@noop {} {\bibfield  {journal} {\bibinfo  {journal} {Biophys.\ J.}\
  }\textbf {\bibinfo {volume} {97}},\ \bibinfo {pages} {83} (\bibinfo {year}
  {2009})}\BibitemShut {NoStop}%
\bibitem [{\citenamefont {Colombelli}\ \emph {et~al.}(2009)\citenamefont
  {Colombelli}, \citenamefont {Besser}, \citenamefont {Kress}, \citenamefont
  {Reynaud}, \citenamefont {Girard}, \citenamefont {Caussinus}, \citenamefont
  {Haselmann}, \citenamefont {Small}, \citenamefont {Schwarz},\ and\
  \citenamefont {Stelzer}}]{Colombelli}%
  \BibitemOpen
  \bibfield  {author} {\bibinfo {author} {\bibfnamefont {J.}~\bibnamefont
  {Colombelli}}, \bibinfo {author} {\bibfnamefont {A.}~\bibnamefont {Besser}},
  \bibinfo {author} {\bibfnamefont {H.}~\bibnamefont {Kress}}, \bibinfo
  {author} {\bibfnamefont {E.~G.}\ \bibnamefont {Reynaud}}, \bibinfo {author}
  {\bibfnamefont {P.}~\bibnamefont {Girard}}, \bibinfo {author} {\bibfnamefont
  {E.}~\bibnamefont {Caussinus}}, \bibinfo {author} {\bibfnamefont
  {U.}~\bibnamefont {Haselmann}}, \bibinfo {author} {\bibfnamefont {J.~V.}\
  \bibnamefont {Small}}, \bibinfo {author} {\bibfnamefont {U.~S.}\ \bibnamefont
  {Schwarz}}, \ and\ \bibinfo {author} {\bibfnamefont {E.~H.~K.}\ \bibnamefont
  {Stelzer}},\ }\href@noop {} {\bibfield  {journal} {\bibinfo  {journal} {J.\
  Cell Sci.}\ }\textbf {\bibinfo {volume} {122}},\ \bibinfo {pages} {1665}
  (\bibinfo {year} {2009})}\BibitemShut {NoStop}%
\bibitem [{\citenamefont {Lu}\ \emph {et~al.}(2008)\citenamefont {Lu},
  \citenamefont {Oswald}, \citenamefont {Ngu},\ and\ \citenamefont
  {Yin}}]{Lu2008}%
  \BibitemOpen
  \bibfield  {author} {\bibinfo {author} {\bibfnamefont {L.}~\bibnamefont
  {Lu}}, \bibinfo {author} {\bibfnamefont {S.~J.}\ \bibnamefont {Oswald}},
  \bibinfo {author} {\bibfnamefont {H.}~\bibnamefont {Ngu}}, \ and\ \bibinfo
  {author} {\bibfnamefont {F.~C.-P.}\ \bibnamefont {Yin}},\ }\href@noop {}
  {\bibfield  {journal} {\bibinfo  {journal} {Biophys.\ J.}\ }\textbf {\bibinfo
  {volume} {95}},\ \bibinfo {pages} {6060} (\bibinfo {year}
  {2008})}\BibitemShut {NoStop}%
\bibitem [{\citenamefont {Hirata}\ \emph {et~al.}(2008)\citenamefont {Hirata},
  \citenamefont {Tatsumi},\ and\ \citenamefont {Sokabe}}]{Hirata2008}%
  \BibitemOpen
  \bibfield  {author} {\bibinfo {author} {\bibfnamefont {H.}~\bibnamefont
  {Hirata}}, \bibinfo {author} {\bibfnamefont {H.}~\bibnamefont {Tatsumi}}, \
  and\ \bibinfo {author} {\bibfnamefont {M.}~\bibnamefont {Sokabe}},\
  }\href@noop {} {\bibfield  {journal} {\bibinfo  {journal} {J.\ Cell Sci.}\
  }\textbf {\bibinfo {volume} {121}},\ \bibinfo {pages} {2795} (\bibinfo {year}
  {2008})}\BibitemShut {NoStop}%
\bibitem [{\citenamefont {Yoshigi}\ \emph {et~al.}(2005)\citenamefont
  {Yoshigi}, \citenamefont {Hoffman}, \citenamefont {Jensen}, \citenamefont
  {Yost},\ and\ \citenamefont {Beckerle}}]{Yoshigi2005}%
  \BibitemOpen
  \bibfield  {author} {\bibinfo {author} {\bibfnamefont {M.}~\bibnamefont
  {Yoshigi}}, \bibinfo {author} {\bibfnamefont {L.~M.}\ \bibnamefont
  {Hoffman}}, \bibinfo {author} {\bibfnamefont {C.~C.}\ \bibnamefont {Jensen}},
  \bibinfo {author} {\bibfnamefont {H.~J.}\ \bibnamefont {Yost}}, \ and\
  \bibinfo {author} {\bibfnamefont {M.~C.}\ \bibnamefont {Beckerle}},\
  }\href@noop {} {\bibfield  {journal} {\bibinfo  {journal} {J.\ Cell Biol.}\
  }\textbf {\bibinfo {volume} {171}},\ \bibinfo {pages} {209} (\bibinfo {year}
  {2005})}\BibitemShut {NoStop}%
\bibitem [{\citenamefont {Burridge}\ and\ \citenamefont
  {Wittchen}(2013)}]{Burridge2013}%
  \BibitemOpen
  \bibfield  {author} {\bibinfo {author} {\bibfnamefont {K.}~\bibnamefont
  {Burridge}}\ and\ \bibinfo {author} {\bibfnamefont {E.~S.}\ \bibnamefont
  {Wittchen}},\ }\href@noop {} {\bibfield  {journal} {\bibinfo  {journal} {J.\
  Cell Biol.}\ }\textbf {\bibinfo {volume} {200}},\ \bibinfo {pages} {9}
  (\bibinfo {year} {2013})}\BibitemShut {NoStop}%
\bibitem [{\citenamefont {Deguchi}\ \emph {et~al.}(2006)\citenamefont
  {Deguchi}, \citenamefont {Ohashi},\ and\ \citenamefont {Sato}}]{Deguchi2006}%
  \BibitemOpen
  \bibfield  {author} {\bibinfo {author} {\bibfnamefont {S.}~\bibnamefont
  {Deguchi}}, \bibinfo {author} {\bibfnamefont {T.}~\bibnamefont {Ohashi}}, \
  and\ \bibinfo {author} {\bibfnamefont {M.}~\bibnamefont {Sato}},\ }\href@noop
  {} {\bibfield  {journal} {\bibinfo  {journal} {J.\ Biomech.}\ }\textbf
  {\bibinfo {volume} {39}},\ \bibinfo {pages} {2603} (\bibinfo {year}
  {2006})}\BibitemShut {NoStop}%
\bibitem [{\citenamefont {Kumar}\ \emph {et~al.}(2006)\citenamefont {Kumar},
  \citenamefont {Maxwell}, \citenamefont {Heisterkamp}, \citenamefont {Polte},
  \citenamefont {Lele}, \citenamefont {Salanga}, \citenamefont {Mazur},\ and\
  \citenamefont {Ingber}}]{Kumar2006}%
  \BibitemOpen
  \bibfield  {author} {\bibinfo {author} {\bibfnamefont {S.}~\bibnamefont
  {Kumar}}, \bibinfo {author} {\bibfnamefont {I.~Z.}\ \bibnamefont {Maxwell}},
  \bibinfo {author} {\bibfnamefont {A.}~\bibnamefont {Heisterkamp}}, \bibinfo
  {author} {\bibfnamefont {T.~R.}\ \bibnamefont {Polte}}, \bibinfo {author}
  {\bibfnamefont {T.~P.}\ \bibnamefont {Lele}}, \bibinfo {author}
  {\bibfnamefont {M.}~\bibnamefont {Salanga}}, \bibinfo {author} {\bibfnamefont
  {E.}~\bibnamefont {Mazur}}, \ and\ \bibinfo {author} {\bibfnamefont {D.~E.}\
  \bibnamefont {Ingber}},\ }\href@noop {} {\bibfield  {journal} {\bibinfo
  {journal} {Biophys.\ J.}\ }\textbf {\bibinfo {volume} {90}},\ \bibinfo
  {pages} {3762} (\bibinfo {year} {2006})}\BibitemShut {NoStop}%
\bibitem [{\citenamefont {Machida}\ \emph {et~al.}(2010)\citenamefont
  {Machida}, \citenamefont {Watanabe-Nakayama}, \citenamefont {Harada},
  \citenamefont {Afrin}, \citenamefont {Nakayama},\ and\ \citenamefont
  {Ikai}}]{Machida2010}%
  \BibitemOpen
  \bibfield  {author} {\bibinfo {author} {\bibfnamefont {S.}~\bibnamefont
  {Machida}}, \bibinfo {author} {\bibfnamefont {T.}~\bibnamefont
  {Watanabe-Nakayama}}, \bibinfo {author} {\bibfnamefont {I.}~\bibnamefont
  {Harada}}, \bibinfo {author} {\bibfnamefont {R.}~\bibnamefont {Afrin}},
  \bibinfo {author} {\bibfnamefont {T.}~\bibnamefont {Nakayama}}, \ and\
  \bibinfo {author} {\bibfnamefont {A.}~\bibnamefont {Ikai}},\ }\href@noop {}
  {\bibfield  {journal} {\bibinfo  {journal} {Nanotechnology}\ }\textbf
  {\bibinfo {volume} {21}},\ \bibinfo {pages} {385102} (\bibinfo {year}
  {2010})}\BibitemShut {NoStop}%
\bibitem [{\citenamefont {Eyckmans}\ \emph {et~al.}(2011)\citenamefont
  {Eyckmans}, \citenamefont {Boudou}, \citenamefont {Yu},\ and\ \citenamefont
  {Chen}}]{Eyckmans11}%
  \BibitemOpen
  \bibfield  {author} {\bibinfo {author} {\bibfnamefont {J.}~\bibnamefont
  {Eyckmans}}, \bibinfo {author} {\bibfnamefont {T.}~\bibnamefont {Boudou}},
  \bibinfo {author} {\bibfnamefont {X.}~\bibnamefont {Yu}}, \ and\ \bibinfo
  {author} {\bibfnamefont {C.~S.}\ \bibnamefont {Chen}},\ }\href@noop {}
  {\bibfield  {journal} {\bibinfo  {journal} {Dev.\ Cell}\ }\textbf {\bibinfo
  {volume} {21}},\ \bibinfo {pages} {35} (\bibinfo {year} {2011})}\BibitemShut
  {NoStop}%
\bibitem [{\citenamefont {Hoffman}\ \emph {et~al.}(2012)\citenamefont
  {Hoffman}, \citenamefont {Jensen}, \citenamefont {Chaturvedi}, \citenamefont
  {Yoshigi},\ and\ \citenamefont {Beckerle}}]{Hoffm12stress-fibers}%
  \BibitemOpen
  \bibfield  {author} {\bibinfo {author} {\bibfnamefont {L.~M.}\ \bibnamefont
  {Hoffman}}, \bibinfo {author} {\bibfnamefont {C.~C.}\ \bibnamefont {Jensen}},
  \bibinfo {author} {\bibfnamefont {A.}~\bibnamefont {Chaturvedi}}, \bibinfo
  {author} {\bibfnamefont {M.}~\bibnamefont {Yoshigi}}, \ and\ \bibinfo
  {author} {\bibfnamefont {M.~C.}\ \bibnamefont {Beckerle}},\ }\href@noop {}
  {\bibfield  {journal} {\bibinfo  {journal} {Mol.\ Biol.\ Cell}\ }\textbf
  {\bibinfo {volume} {23}},\ \bibinfo {pages} {1846} (\bibinfo {year}
  {2012})}\BibitemShut {NoStop}%
\bibitem [{\citenamefont {Yasukuni}\ \emph {et~al.}(2007)\citenamefont
  {Yasukuni}, \citenamefont {Spitz}, \citenamefont {Meallet-Renault},
  \citenamefont {Negishi}, \citenamefont {Tada}, \citenamefont {Hosokawa},
  \citenamefont {Asahi}, \citenamefont {Shukunami}, \citenamefont {Hiraki},\
  and\ \citenamefont {Masuhara}}]{Yasukuni2007}%
  \BibitemOpen
  \bibfield  {author} {\bibinfo {author} {\bibfnamefont {R.}~\bibnamefont
  {Yasukuni}}, \bibinfo {author} {\bibfnamefont {J.-A.}\ \bibnamefont {Spitz}},
  \bibinfo {author} {\bibfnamefont {R.}~\bibnamefont {Meallet-Renault}},
  \bibinfo {author} {\bibfnamefont {T.}~\bibnamefont {Negishi}}, \bibinfo
  {author} {\bibfnamefont {T.}~\bibnamefont {Tada}}, \bibinfo {author}
  {\bibfnamefont {Y.}~\bibnamefont {Hosokawa}}, \bibinfo {author}
  {\bibfnamefont {T.}~\bibnamefont {Asahi}}, \bibinfo {author} {\bibfnamefont
  {C.}~\bibnamefont {Shukunami}}, \bibinfo {author} {\bibfnamefont
  {Y.}~\bibnamefont {Hiraki}}, \ and\ \bibinfo {author} {\bibfnamefont
  {H.}~\bibnamefont {Masuhara}},\ }\href@noop {} {\bibfield  {journal}
  {\bibinfo  {journal} {Appl.\ Surf.\ Sci.}\ }\textbf {\bibinfo {volume}
  {253}},\ \bibinfo {pages} {6416} (\bibinfo {year} {2007})}\BibitemShut
  {NoStop}%
\bibitem [{\citenamefont {Helfrich}(1995)}]{HelfrichBook}%
  \BibitemOpen
  \bibfield  {author} {\bibinfo {author} {\bibfnamefont {W.}~\bibnamefont
  {Helfrich}},\ }\href@noop {} {\emph {\bibinfo {title} {Handbook of
  {B}iological {P}hysics {V}ol.~1}}},\ edited by\ \bibinfo {editor}
  {\bibfnamefont {R.}~\bibnamefont {Lipowsky}}\ and\ \bibinfo {editor}
  {\bibfnamefont {E.}~\bibnamefont {Sackmann}}\ (\bibinfo  {publisher}
  {Elsevier Science B.~V.},\ \bibinfo {year} {1995})\BibitemShut {NoStop}%
\bibitem [{\citenamefont {Seifert}(1995)}]{SeifertPRL95}%
  \BibitemOpen
  \bibfield  {author} {\bibinfo {author} {\bibfnamefont {U.}~\bibnamefont
  {Seifert}},\ }\href@noop {} {\bibfield  {journal} {\bibinfo  {journal}
  {Phys.\ Rev.\ Lett.}\ }\textbf {\bibinfo {volume} {74}},\ \bibinfo {pages}
  {5060} (\bibinfo {year} {1995})}\BibitemShut {NoStop}%
\bibitem [{\citenamefont {Sackmann}\ and\ \citenamefont
  {Bruinsma}(2002)}]{SackmBruinsma02}%
  \BibitemOpen
  \bibfield  {author} {\bibinfo {author} {\bibfnamefont {E.}~\bibnamefont
  {Sackmann}}\ and\ \bibinfo {author} {\bibfnamefont {R.~F.}\ \bibnamefont
  {Bruinsma}},\ }\href@noop {} {\bibfield  {journal} {\bibinfo  {journal}
  {ChemPhysChem}\ }\textbf {\bibinfo {volume} {3}},\ \bibinfo {pages} {262}
  (\bibinfo {year} {2002})}\BibitemShut {NoStop}%
\bibitem [{\citenamefont {Sengupta}\ and\ \citenamefont
  {Limozin}(2010)}]{SenguptaPRL10}%
  \BibitemOpen
  \bibfield  {author} {\bibinfo {author} {\bibfnamefont {K.}~\bibnamefont
  {Sengupta}}\ and\ \bibinfo {author} {\bibfnamefont {L.}~\bibnamefont
  {Limozin}},\ }\href@noop {} {\bibfield  {journal} {\bibinfo  {journal}
  {Phys.\ Rev.\ Lett.}\ }\textbf {\bibinfo {volume} {104}},\ \bibinfo {pages}
  {088101} (\bibinfo {year} {2010})}\BibitemShut {NoStop}%
\bibitem [{\citenamefont {Tamada}\ \emph {et~al.}(2004)\citenamefont {Tamada},
  \citenamefont {Sheetz},\ and\ \citenamefont {Sawada}}]{Tamada04}%
  \BibitemOpen
  \bibfield  {author} {\bibinfo {author} {\bibfnamefont {M.}~\bibnamefont
  {Tamada}}, \bibinfo {author} {\bibfnamefont {M.~P.}\ \bibnamefont {Sheetz}},
  \ and\ \bibinfo {author} {\bibfnamefont {Y.}~\bibnamefont {Sawada}},\
  }\href@noop {} {\bibfield  {journal} {\bibinfo  {journal} {Dev.\ Cell}\
  }\textbf {\bibinfo {volume} {7}},\ \bibinfo {pages} {709} (\bibinfo {year}
  {2004})}\BibitemShut {NoStop}%
\bibitem [{\citenamefont {Wojtowicz}\ \emph {et~al.}(2010)\citenamefont
  {Wojtowicz}, \citenamefont {Babu}, \citenamefont {Li}, \citenamefont {Gretz},
  \citenamefont {Hecker},\ and\ \citenamefont {Cattaruzza}}]{Wojtowicz10}%
  \BibitemOpen
  \bibfield  {author} {\bibinfo {author} {\bibfnamefont {A.}~\bibnamefont
  {Wojtowicz}}, \bibinfo {author} {\bibfnamefont {S.~S.}\ \bibnamefont {Babu}},
  \bibinfo {author} {\bibfnamefont {L.}~\bibnamefont {Li}}, \bibinfo {author}
  {\bibfnamefont {N.}~\bibnamefont {Gretz}}, \bibinfo {author} {\bibfnamefont
  {M.}~\bibnamefont {Hecker}}, \ and\ \bibinfo {author} {\bibfnamefont
  {M.}~\bibnamefont {Cattaruzza}},\ }\href@noop {} {\bibfield  {journal}
  {\bibinfo  {journal} {Circ.\ Res.}\ }\textbf {\bibinfo {volume} {107}},\
  \bibinfo {pages} {898} (\bibinfo {year} {2010})}\BibitemShut {NoStop}%
\bibitem [{\citenamefont {Yang}\ \emph {et~al.}(2013)\citenamefont {Yang},
  \citenamefont {Bai}, \citenamefont {Klug}, \citenamefont {Levine},\ and\
  \citenamefont {Valentine}}]{YangMicrorheo13}%
  \BibitemOpen
  \bibfield  {author} {\bibinfo {author} {\bibfnamefont {Y.}~\bibnamefont
  {Yang}}, \bibinfo {author} {\bibfnamefont {M.}~\bibnamefont {Bai}}, \bibinfo
  {author} {\bibfnamefont {W.~S.}\ \bibnamefont {Klug}}, \bibinfo {author}
  {\bibfnamefont {A.~J.}\ \bibnamefont {Levine}}, \ and\ \bibinfo {author}
  {\bibfnamefont {M.~T.}\ \bibnamefont {Valentine}},\ }\href@noop {} {\bibfield
   {journal} {\bibinfo  {journal} {Soft Matter}\ }\textbf {\bibinfo {volume}
  {9}},\ \bibinfo {pages} {383} (\bibinfo {year} {2013})}\BibitemShut {NoStop}%
\bibitem [{\citenamefont {Heussinger}(2011)}]{Claus11coop-unbind-bundle}%
  \BibitemOpen
  \bibfield  {author} {\bibinfo {author} {\bibfnamefont {C.}~\bibnamefont
  {Heussinger}},\ }\href@noop {} {\bibfield  {journal} {\bibinfo  {journal}
  {Phys.\ Rev.\ E}\ }\textbf {\bibinfo {volume} {83}},\ \bibinfo {pages}
  {050902(R)} (\bibinfo {year} {2011})}\BibitemShut {NoStop}%
\bibitem [{\citenamefont {Vink}\ and\ \citenamefont
  {Heussinger}(2012)}]{RichardClaus12}%
  \BibitemOpen
  \bibfield  {author} {\bibinfo {author} {\bibfnamefont {R.~L.~C.}\
  \bibnamefont {Vink}}\ and\ \bibinfo {author} {\bibfnamefont {C.}~\bibnamefont
  {Heussinger}},\ }\href@noop {} {\bibfield  {journal} {\bibinfo  {journal}
  {J.\ Chem.\ Phys.}\ }\textbf {\bibinfo {volume} {136}},\ \bibinfo {pages}
  {035102} (\bibinfo {year} {2012})}\BibitemShut {NoStop}%
\bibitem [{\citenamefont {Gross}\ \emph {et~al.}(2011)\citenamefont {Gross},
  \citenamefont {Laurens}, \citenamefont {Oddershede}, \citenamefont
  {Bockelmann}, \citenamefont {Petermann},\ and\ \citenamefont
  {Wuite}}]{GrossDNAmeltNatPhys11}%
  \BibitemOpen
  \bibfield  {author} {\bibinfo {author} {\bibfnamefont {P.}~\bibnamefont
  {Gross}}, \bibinfo {author} {\bibfnamefont {N.}~\bibnamefont {Laurens}},
  \bibinfo {author} {\bibfnamefont {L.~B.}\ \bibnamefont {Oddershede}},
  \bibinfo {author} {\bibfnamefont {U.}~\bibnamefont {Bockelmann}}, \bibinfo
  {author} {\bibfnamefont {E.~J.~G.}\ \bibnamefont {Petermann}}, \ and\
  \bibinfo {author} {\bibfnamefont {G.~J.~L.}\ \bibnamefont {Wuite}},\
  }\href@noop {} {\bibfield  {journal} {\bibinfo  {journal} {Nat.\ Phys.}\
  }\textbf {\bibinfo {volume} {7}},\ \bibinfo {pages} {731} (\bibinfo {year}
  {2011})}\BibitemShut {NoStop}%
\bibitem [{\citenamefont {Marenduzzo}\ \emph {et~al.}(2009)\citenamefont
  {Marenduzzo}, \citenamefont {Maritan}, \citenamefont {Orlandini},
  \citenamefont {Seno},\ and\ \citenamefont {Trovato}}]{Marenduzzo09}%
  \BibitemOpen
  \bibfield  {author} {\bibinfo {author} {\bibfnamefont {D.}~\bibnamefont
  {Marenduzzo}}, \bibinfo {author} {\bibfnamefont {A.}~\bibnamefont {Maritan}},
  \bibinfo {author} {\bibfnamefont {E.}~\bibnamefont {Orlandini}}, \bibinfo
  {author} {\bibfnamefont {F.}~\bibnamefont {Seno}}, \ and\ \bibinfo {author}
  {\bibfnamefont {A.}~\bibnamefont {Trovato}},\ }\href@noop {} {\bibfield
  {journal} {\bibinfo  {journal} {J.\ Stat.\ Mech.}\ }\textbf {\bibinfo
  {volume} {2009}},\ \bibinfo {pages} {L04001} (\bibinfo {year}
  {2009})}\BibitemShut {NoStop}%
\bibitem [{\citenamefont {Theodorakopoulos}\ and\ \citenamefont
  {Peyrard}(2012)}]{TheoPeyr12}%
  \BibitemOpen
  \bibfield  {author} {\bibinfo {author} {\bibfnamefont {N.}~\bibnamefont
  {Theodorakopoulos}}\ and\ \bibinfo {author} {\bibfnamefont {M.}~\bibnamefont
  {Peyrard}},\ }\href@noop {} {\bibfield  {journal} {\bibinfo  {journal}
  {Phys.\ Rev.\ Lett.}\ }\textbf {\bibinfo {volume} {108}},\ \bibinfo {pages}
  {078104} (\bibinfo {year} {2012})}\BibitemShut {NoStop}%
\bibitem [{\citenamefont {Marko}\ and\ \citenamefont
  {Siggia}(1995)}]{MarkoSiggia1995}%
  \BibitemOpen
  \bibfield  {author} {\bibinfo {author} {\bibfnamefont {J.~F.}\ \bibnamefont
  {Marko}}\ and\ \bibinfo {author} {\bibfnamefont {E.~D.}\ \bibnamefont
  {Siggia}},\ }\href@noop {} {\bibfield  {journal} {\bibinfo  {journal}
  {Macromolecules}\ }\textbf {\bibinfo {volume} {28}},\ \bibinfo {pages} {8759}
  (\bibinfo {year} {1995})}\BibitemShut {NoStop}%
\bibitem [{\citenamefont {von~der Heydt}\ \emph {et~al.}(2013)\citenamefont
  {von~der Heydt}, \citenamefont {Wilkin}, \citenamefont {Benetatos},\ and\
  \citenamefont {Zippelius}}]{AliceX-linkedWLC13}%
  \BibitemOpen
  \bibfield  {author} {\bibinfo {author} {\bibfnamefont {A.}~\bibnamefont
  {von~der Heydt}}, \bibinfo {author} {\bibfnamefont {D.}~\bibnamefont
  {Wilkin}}, \bibinfo {author} {\bibfnamefont {P.}~\bibnamefont {Benetatos}}, \
  and\ \bibinfo {author} {\bibfnamefont {A.}~\bibnamefont {Zippelius}},\
  }\href@noop {} {\bibfield  {journal} {\bibinfo  {journal} {Phys.\ Rev.\ E}\
  }\textbf {\bibinfo {volume} {88}},\ \bibinfo {pages} {032701} (\bibinfo
  {year} {2013})}\BibitemShut {NoStop}%
\bibitem [{\citenamefont {Nelson}(2002)}]{NelsonBook}%
  \BibitemOpen
  \bibfield  {author} {\bibinfo {author} {\bibfnamefont {D.~R.}\ \bibnamefont
  {Nelson}},\ }\enquote {\bibinfo {title} {Defects and {G}eometry in
  {C}ondensed {M}atter {P}hysics},}\ \ (\bibinfo  {publisher} {Cambridge Univ.\
  Press},\ \bibinfo {address} {Cambridge},\ \bibinfo {year} {2002})\
  Chap.~\bibinfo {chapter} {7}, p.\ \bibinfo {pages} {251}\BibitemShut
  {NoStop}%
\bibitem [{\citenamefont {Kierfeld}\ \emph {et~al.}(2005)\citenamefont
  {Kierfeld}, \citenamefont {K\"uhne},\ and\ \citenamefont
  {Lipowsky}}]{Kierfeld05}%
  \BibitemOpen
  \bibfield  {author} {\bibinfo {author} {\bibfnamefont {J.}~\bibnamefont
  {Kierfeld}}, \bibinfo {author} {\bibfnamefont {T.}~\bibnamefont {K\"uhne}}, \
  and\ \bibinfo {author} {\bibfnamefont {R.}~\bibnamefont {Lipowsky}},\
  }\href@noop {} {\bibfield  {journal} {\bibinfo  {journal} {Phys.\ Rev.\
  Lett.}\ }\textbf {\bibinfo {volume} {95}},\ \bibinfo {pages} {038102}
  (\bibinfo {year} {2005})}\BibitemShut {NoStop}%
\bibitem [{\citenamefont {Benetatos}\ and\ \citenamefont
  {Terentjev}(2010)}]{PBEMT10R}%
  \BibitemOpen
  \bibfield  {author} {\bibinfo {author} {\bibfnamefont {P.}~\bibnamefont
  {Benetatos}}\ and\ \bibinfo {author} {\bibfnamefont {E.~M.}\ \bibnamefont
  {Terentjev}},\ }\href@noop {} {\bibfield  {journal} {\bibinfo  {journal}
  {Phys.\ Rev.\ E}\ }\textbf {\bibinfo {volume} {82}},\ \bibinfo {pages}
  {050802(R)} (\bibinfo {year} {2010})}\BibitemShut {NoStop}%
\bibitem [{\citenamefont {Feynman}(1972)}]{Feynman}%
  \BibitemOpen
  \bibfield  {author} {\bibinfo {author} {\bibfnamefont {R.~P.}\ \bibnamefont
  {Feynman}},\ }\href@noop {} {\emph {\bibinfo {title} {Statistical
  Mechanics}}}\ (\bibinfo  {publisher} {Addison-Wesley},\ \bibinfo {address}
  {Menlo Park},\ \bibinfo {year} {1972})\BibitemShut {NoStop}%
\bibitem [{\citenamefont {Landau}\ and\ \citenamefont
  {Lifshitz}(1958)}]{Landau}%
  \BibitemOpen
  \bibfield  {author} {\bibinfo {author} {\bibfnamefont {L.~D.}\ \bibnamefont
  {Landau}}\ and\ \bibinfo {author} {\bibfnamefont {E.~M.}\ \bibnamefont
  {Lifshitz}},\ }\href@noop {} {\emph {\bibinfo {title} {Quantum
  {M}echanics}}}\ (\bibinfo  {publisher} {Pergamon Press},\ \bibinfo {address}
  {London},\ \bibinfo {year} {1958})\BibitemShut {NoStop}%
\bibitem [{\citenamefont {Benetatos}\ \emph {et~al.}(2012)\citenamefont
  {Benetatos}, \citenamefont {Ulrich},\ and\ \citenamefont
  {Zippelius}}]{PanayotisStephanAnnette12}%
  \BibitemOpen
  \bibfield  {author} {\bibinfo {author} {\bibfnamefont {P.}~\bibnamefont
  {Benetatos}}, \bibinfo {author} {\bibfnamefont {S.}~\bibnamefont {Ulrich}}, \
  and\ \bibinfo {author} {\bibfnamefont {A.}~\bibnamefont {Zippelius}},\
  }\href@noop {} {\bibfield  {journal} {\bibinfo  {journal} {New J.\ Phys.}\
  }\textbf {\bibinfo {volume} {14}},\ \bibinfo {pages} {115011} (\bibinfo
  {year} {2012})}\BibitemShut {NoStop}%
\bibitem [{\citenamefont {Heussinger}\ \emph {et~al.}(2007)\citenamefont
  {Heussinger}, \citenamefont {Bathe},\ and\ \citenamefont
  {Frey}}]{Claus07PRL}%
  \BibitemOpen
  \bibfield  {author} {\bibinfo {author} {\bibfnamefont {C.}~\bibnamefont
  {Heussinger}}, \bibinfo {author} {\bibfnamefont {M.}~\bibnamefont {Bathe}}, \
  and\ \bibinfo {author} {\bibfnamefont {E.}~\bibnamefont {Frey}},\ }\href@noop
  {} {\bibfield  {journal} {\bibinfo  {journal} {Phys.\ Rev.\ Lett.}\ }\textbf
  {\bibinfo {volume} {99}},\ \bibinfo {pages} {048101} (\bibinfo {year}
  {2007})}\BibitemShut {NoStop}%
\bibitem [{\citenamefont {Benetatos}\ and\ \citenamefont
  {Frey}(2003)}]{PBEFrey03}%
  \BibitemOpen
  \bibfield  {author} {\bibinfo {author} {\bibfnamefont {P.}~\bibnamefont
  {Benetatos}}\ and\ \bibinfo {author} {\bibfnamefont {E.}~\bibnamefont
  {Frey}},\ }\href@noop {} {\bibfield  {journal} {\bibinfo  {journal} {Phys.\
  Rev.\ E}\ }\textbf {\bibinfo {volume} {67}},\ \bibinfo {pages} {051108}
  (\bibinfo {year} {2003})}\BibitemShut {NoStop}%
\bibitem [{\citenamefont {Benetatos}\ and\ \citenamefont
  {Zippelius}(2007)}]{PBAZPRL07}%
  \BibitemOpen
  \bibfield  {author} {\bibinfo {author} {\bibfnamefont {P.}~\bibnamefont
  {Benetatos}}\ and\ \bibinfo {author} {\bibfnamefont {A.}~\bibnamefont
  {Zippelius}},\ }\href@noop {} {\bibfield  {journal} {\bibinfo  {journal}
  {Phys.\ Rev.\ Lett.}\ }\textbf {\bibinfo {volume} {99}},\ \bibinfo {pages}
  {198301} (\bibinfo {year} {2007})}\BibitemShut {NoStop}%
\bibitem [{\citenamefont {Liverpool}\ \emph {et~al.}(2009)\citenamefont
  {Liverpool}, \citenamefont {Marchetti}, \citenamefont {Joanny},\ and\
  \citenamefont {Prost}}]{LiverpoolEPL09}%
  \BibitemOpen
  \bibfield  {author} {\bibinfo {author} {\bibfnamefont {T.~B.}\ \bibnamefont
  {Liverpool}}, \bibinfo {author} {\bibfnamefont {M.~C.}\ \bibnamefont
  {Marchetti}}, \bibinfo {author} {\bibfnamefont {J.-F.}\ \bibnamefont
  {Joanny}}, \ and\ \bibinfo {author} {\bibfnamefont {J.}~\bibnamefont
  {Prost}},\ }\href@noop {} {\bibfield  {journal} {\bibinfo  {journal}
  {Europhys.\ Lett.}\ }\textbf {\bibinfo {volume} {85}},\ \bibinfo {pages}
  {18007} (\bibinfo {year} {2009})}\BibitemShut {NoStop}%
\end{thebibliography}%




\end{document}